\documentclass[conference]{IEEEtran}
\IEEEoverridecommandlockouts

\usepackage{amsmath,amssymb,amsfonts}
\usepackage{algorithmic}
\usepackage{graphicx}
\usepackage{textcomp}
\usepackage{xcolor}
\usepackage{cite}
\usepackage{booktabs}
\usepackage{hyperref}
\hypersetup{hidelinks}

\begin{document}

\title{Optimizing Intra-Container Communication with Memory Protection Keys: A Novel Approach to Secure and Efficient Microservice Interaction}

\author{
\IEEEauthorblockN{Fnu Yashu}
\IEEEauthorblockA{
    Stony Brook University\\
    Dept. of Computer Science\\
    Stony Brook, NY, USA \\
    \textnormal{\texttt{yyashu@cs.stonybrook.edu}} 
}
\and
\IEEEauthorblockN{Shubham Malhotra}
\IEEEauthorblockA{
    Rochester Institute of Technology\\
    Dept. of Software Engineering\\
    Rochester, NY, USA \\
    \textnormal{\texttt{shubham.malhotra28@gmail.com}} 
}
\and
\IEEEauthorblockN{Muhammad Saqib}
\IEEEauthorblockA{
    Texas Tech University\\
    Dept. of Computer Science\\
    Lubbock, TX, USA \\
    \textnormal{\texttt{saqibraopk@hotmail.com}} 
}
}

\maketitle
\begin{abstract}
In modern cloud-native applications, microservices are commonly deployed in containerized environments to ensure scalability and flexibility. However, inter-process communication (IPC) between co-located microservices often suffers from significant overhead, especially when traditional networking protocols are employed within containers. This paper introduces a novel approach, MPKLink, leveraging Intel Memory Protection Keys (MPK) to enhance intra-container communication efficiency while ensuring security. By utilizing shared memory with MPK-based access control, we eliminate unnecessary networking latencies, leading to reduced resource consumption and faster response times. We present a comprehensive evaluation of MPKLink, demonstrating its superior performance over conventional methods such as REST and gRPC within microservice architectures. Furthermore, we explore the integration of this approach with existing container orchestration platforms, showcasing its seamless adoption in real-world deployment scenarios. This work provides a transformative solution for developers looking to optimize communication in microservices while maintaining the integrity and security of containerized applications.
\end{abstract}


\section{Introduction}
The adoption of microservices has fundamentally transformed software development by promoting modularity, ease of maintenance, and scalability. These loosely coupled components streamline system architecture but introduce communication overhead, particularly when deployed within the same containerized environment. Conventional communication techniques, including Remote Procedure Calls (RPCs) and REST APIs, often mimic network-based interactions even when microservices reside on the same physical host. This results in inefficient resource utilization due to unnecessary protocol overhead.

To address these inefficiencies, shared memory mechanisms present a compelling alternative, offering direct data exchange between co-located microservices without traversing the network stack. However, existing shared memory implementations often lack sufficient security mechanisms, making them susceptible to unauthorized access and data corruption.

Industry trends and academic research emphasize the shortcomings of traditional IPC methods for co-located microservices. Many deployments continue to rely on REST, gRPC, or Thrift secured through JWT authentication or mutual TLS encryption. While effective in distributed environments, these protocols impose unwarranted complexity and latency in scenarios where microservices share the same host. Consequently, there is a pressing need for optimized communication techniques that balance high performance with robust security measures.

This paper introduces MPKLink, a novel solution leveraging Intel's Memory Protection Keys (MPK) to achieve efficient intra-container communication while maintaining strict access control. MPKLink enables microservices to utilize shared memory with dynamic permission management, reducing computational overhead while preserving security. The subsequent sections provide a detailed analysis of MPKLink’s implementation, performance benefits, and practical integration within modern container orchestration ecosystems. MPKLink distinguishes itself from traditional shared memory implementations and network-based IPC methods (e.g., REST and gRPC) by leveraging Intel Memory Protection Keys (MPK) for dynamic and fine-grained access control. Unlike conventional methods that lack adaptive security measures, MPKLink allows rapid permission updates at the thread level—ensuring robust isolation and significantly reducing protocol overhead. This enhanced security model not only mitigates unauthorized memory access but also minimizes the risks associated with typical attack vectors, such as man-in-the-middle attacks and side-channel exploits, thereby providing a compelling advantage for secure intra-container communications.

\section{Threat Model}
A threat model is assumed in which multiple microservices are executed on the same host machine. Such a configuration is common in modern containerized environments—managed by tools like Docker Compose—where microservices are deployed in isolated containers while sharing the underlying hardware and kernel. To enable efficient communication, shared memory or other inter-process communication (IPC) mechanisms are typically employed via a shared filesystem mount on the host. Although performance benefits are achieved with this approach, potential security vulnerabilities are introduced when one or more microservices become compromised.

In the assumed threat model, it is presumed that an adversary has compromised one or more microservices while not affecting all of them. Additionally, unprivileged access to the host machine may have been obtained through non-root user credentials or a service account. With this level of access, exploitation of shared resources—such as memory or the filesystem—is attempted to breach the confidentiality or integrity of communication between microservices. One prominent attack vector is the man-in-the-middle (MITM) attack, where data exchanged through shared memory regions is intercepted or manipulated. For example, sensitive data might be read from memory, tampered with prior to its intended use, or malicious content may be injected to disrupt system operations.

Even though containerization provides some isolation, the shared use of host resources can inadvertently create pathways for adversaries. Robust protection of communication channels, prevention of unauthorized access, and assurance of data integrity are required to counter these challenges without sacrificing the performance benefits of shared memory-based communication.

\section{Contributions}
A novel system, termed \textbf{MPKLink}, is introduced to enhance inter-microservice communication for co-located microservices by leveraging \textbf{Intel Memory Protection Keys (MPK)}. Previous research that employed MPK predominantly focused on securing memory at rest by ensuring the integrity and isolation of static or infrequently accessed memory regions. In contrast, the concept is extended to memory in transit by MPKLink. Dynamic and efficient communication between microservices is enabled through shared memory regions secured by MPK, thereby addressing both performance and security challenges inherent in traditional IPC mechanisms.

A practical framework is provided that utilizes MPK to facilitate high-performance and trusted communication without necessitating modifications to preexisting microservices. By securely sharing memory protection keys between services, confidentiality and integrity are preserved against adversarial threats as outlined in the threat model. Performance benchmarks against traditional communication protocols have been conducted to demonstrate the advantages of MPKLink, and its isolation guarantees have been analyzed and validated for compatibility with real-world microservices architectures. It is also noted that no external collaborators were involved, the project is not used for another class, and it was initiated during the current quarter.

\section{Background}
\subsection{gRPC}
gRPC is characterized as a high-performance, open-source framework that facilitates communication between services in microservice architectures. For instance, when a Ruby service initiates a gRPC call to an Android-Java service, a client stub generated during the build process is invoked. This client stub is responsible for encoding the data into Protocol Buffers—a language-neutral, platform-neutral binary format—which is then transmitted over the network to the target service. The data is sent as a stream of HTTP/2 data frames, which provide multiplexing, flow control, and enhanced performance relative to older HTTP protocols. Due to the binary encoding and network optimizations, gRPC is often observed to be up to five times faster than transmitting JSON over HTTP, particularly in high-throughput scenarios.

Upon receipt, the target service decodes the Protocol Buffers-formatted packets and invokes the corresponding application logic. The processed result is then re-encoded into Protocol Buffers format, transmitted through the transport layer, and returned to the client. The compactness and efficiency of Protocol Buffers reduce the overhead compared to other serialization methods, such as JSON, thereby enabling rapid and reliable communication between microservices.

\subsection{Inter Process Communication (IPC)}
When microservices communicate on the same machine, various IPC mechanisms are typically employed. IPC refers to the techniques that allow processes within the same operating system to exchange data and coordinate activities. These techniques are essential for tasks such as data sharing, synchronization, and signaling. UNIX systems provide several forms of IPC, including named pipes, UNIX domain sockets, and shared memory.

\subsubsection{Named Pipes}
Named pipes, also known as FIFOs (First In, First Out), are used as an IPC mechanism that permits data to flow unidirectionally between processes. In contrast to anonymous pipes—limited to parent-child relationships—named pipes are represented as special files in the filesystem and are accessible by unrelated processes. The creation of a named pipe is performed using the \textit{mkfifo} library function, which generates a pipe within the filesystem. Processes interact with the named pipe using standard file operations such as \textit{open}, \textit{read}, and \textit{write}, with one process writing data and another reading it.

A limitation of named pipes is that data flows only in one direction per pipe. Consequently, bidirectional communication between two microservices requires the creation of two separate named pipes—one for sending and one for receiving data. This setup introduces additional complexity, as the correct management and synchronization of multiple pipes become necessary. Moreover, since named pipes typically operate in blocking mode, a process may be stalled if the corresponding reader or writer is inactive, necessitating careful coordination to avoid deadlocks or delays.

\subsubsection{UNIX Domain Sockets}
Unix domain sockets (UDS) serve as an IPC mechanism that enables efficient data exchange between processes on the same machine. Unlike network sockets that utilize protocols such as TCP/IP, UDS operate within the kernel using the \textbf{AF\_UNIX} address family. This characteristic renders them faster and more lightweight than network-based communication methods. UDS support both stream-oriented (akin to TCP) and datagram-oriented (similar to UDP) communications, offering flexibility for different use cases. Additionally, as UDS operate solely on the local machine, they inherently provide enhanced security and reliability compared to network communication.

The implementation of Unix domain sockets involves the socket API with the \textbf{AF\_UNIX} address family. Typically, a socket is created (for example, using \textit{socket(AF\_UNIX, SOCK\_STREAM, 0)}) for stream-oriented communication. One process binds the socket to a filesystem file using \textit{bind()} and then listens for connections with \textit{listen()}, while another process connects using \textit{connect()}. Once established, data is exchanged bidirectionally using standard functions such as \textit{send()} and \textit{recv()}. For instance, a server might create a socket file at \texttt{/tmp/mysocket}, and a client would use the same path to establish a connection.

A significant advantage of Unix domain sockets over named pipes is the support for bidirectional communication, which eliminates the need for multiple pipes. However, proper management of the socket file is required, including cleanup procedures to avoid stale files. Furthermore, UDS may be subject to kernel-imposed limitations, such as buffer sizes, which might require tuning for performance-critical applications. Although highly efficient for local communication, UDS are confined to processes on the same host and are not suitable for distributed architectures.

\subsubsection{Shared Memory}
Shared memory is recognized as a high-performance IPC technique that allows multiple processes on the same system to access a common memory region directly. This method eliminates the overhead associated with copying data between processes and is particularly beneficial for applications that require frequent exchange of large data volumes. However, due to the concurrent access to a common memory region, additional synchronization mechanisms must be implemented to ensure data consistency and prevent race conditions.

Shared memory can be implemented using the System V \textit{shmget} API or via memory-mapped files with \textit{mmap}. With \textit{shmget}, a shared memory segment is created using a unique key and attached by processes using \textit{shmat}, allowing direct read and write operations. Alternatively, \textit{mmap} can be used to map a file (or employ \textit{MAP\_ANONYMOUS} for unnamed memory) into the address space of multiple processes. For example, a shared region may be created by mapping a file located in \texttt{/dev/shm}, thereby enabling rapid communication without the need for intermediate data transfers.

One key attribute of shared memory is its non-blocking nature, meaning that it does not inherently manage synchronization. To prevent race conditions, explicit mechanisms such as locks, mutexes, or semaphores must be employed. In some implementations, separate shared memory regions are allocated for reading and writing to reduce contention, although this increases complexity. Proper resource management is also necessary to ensure that shared memory segments are explicitly cleaned up after use, preventing memory leaks or the persistence of stale data. Despite these challenges, shared memory remains a powerful IPC tool for achieving high-speed communication between processes on the same host.

\subsection{Problems with Shared Memory}
The utilization of shared memory for IPC is accompanied by several challenges that can affect both the security and performance of applications. These issues primarily arise from the absence of built-in mechanisms for fine-grained access control and synchronization, requiring manual implementation of these functionalities. Although high-speed communication is achieved through direct memory access, the complexity and potential for error in managing shared memory can lead to vulnerabilities and performance degradation, particularly in systems that demand strong security and reliability.

\subsubsection{Lack of Fine-Grained Access Control}
Shared memory typically operates with broad access permissions, such that any process with knowledge of the shared memory key may read from or write to the region. Since these keys can often be enumerated or guessed, the risk to confidentiality and integrity is increased. Unauthorized processes may access sensitive data, inject malicious content, or corrupt the memory if additional safeguards—such as enforced policies or key obfuscation—are not applied.

\subsubsection{Synchronization Overhead}
Unlike IPC mechanisms that inherently handle coordination, shared memory requires explicit synchronization to avoid race conditions when multiple processes attempt concurrent access. Mechanisms such as semaphores, mutexes, or spinlocks are typically employed, but these introduce additional overhead. Processes must acquire and release locks before accessing the shared region, and poor synchronization management can result in deadlocks, priority inversion, or performance bottlenecks in high-concurrency scenarios.

\subsubsection{Copying Overhead}
To alleviate synchronization issues, multiple shared memory buffers may be allocated, with specific buffers designated for reading and others for writing. Although this approach minimizes contention by allowing independent access, it incurs additional copying overhead as data is duplicated across regions to maintain consistency. This trade-off can offset some of the inherent performance advantages of shared memory, especially in environments where memory bandwidth is a critical resource.

\subsection{Intel Memory Protection Keys (MPK)}
Intel Memory Protection Keys (MPK) are designed as a user-space mechanism for managing memory permissions more efficiently than traditional methods. Memory pages are tagged with protection keys that specify access rights such as read, write, or execute. These keys are stored in the \textbf{PKRU} register, which is thread-local, thereby allowing different threads within the same process to have distinct access rights for separate memory regions.

One of the primary advantages of MPK is that memory permissions can be modified without resorting to costly system calls like \textit{mprotect}. Once memory pages have been associated with protection keys, the PKRU register can be updated to change access rights—a process that is substantially less expensive than invoking \textit{mprotect}. Because the PKRU register is thread-local, its value must be synchronized between threads or processes to maintain consistent memory access rights. This synchronization is efficiently achieved by storing the PKRU register value in a small shared memory region that is mapped between the participating entities. In this manner, identical memory protection settings are maintained with minimal synchronization overhead, thereby providing a flexible and rapid mechanism for managing memory permissions.
\section{System Design}
Intel MPKLink is engineered as a prospective augmentation to existing microservices orchestration platforms, such as Docker Compose, with the objective of enhancing the security and efficiency of communications among co-located microservices. In a conventional microservices deployment, each service communicates over a virtualized network employing standard protocols like gRPC or REST APIs. MPKLink operates concurrently with these established methods by providing an additional security layer and performance enhancements. By exploiting Intel Memory Protection Keys (MPK) to regulate memory access privileges, MPKLink ensures both high efficiency and stringent isolation among services.

MPKLink assumes the responsibility of instantiating each microservice as an independent thread, thereby enabling communication via shared memory regions safeguarded by MPK. During the initialization of each microservice, memory protection keys are assigned to the respective memory regions utilized for inter-process communication. The system meticulously synchronizes these protection keys across threads to ensure that only designated threads have access to specific memory regions, thereby preventing unauthorized access and preserving data integrity. In addition to managing memory access rights, MPKLink conducts digital signature verifications on each microservice to guarantee that only authenticated and trusted services participate in the communication. This verification process provides an extra layer of defense, ensuring that malicious or unverified microservices are incapable of tampering with protected memory regions.

To further mitigate the risk of exfiltration or misuse of synchronized memory protection keys, each microservice is registered with a unique public-private key pair. MPKLink functions in the role of a dedicated Certificate Authority (CA) by managing these keys and ensuring that only authorized services may exchange information via shared memory. The public key is employed to verify the authenticity of the microservices, while the private key secures access to the service's designated memory regions. This architectural design effectively prevents adversaries from hijacking memory protection keys, thereby offering a robust defense against attacks aimed at extracting or manipulating sensitive data.

In order to minimize extensive modifications to preexisting microservices, MPKLink integrates a modified gRPC or protocol buffer compiler that automatically signs communication packets. As a consequence, every message transmitted between microservices is digitally signed and subsequently verified, thus enabling the seamless incorporation of secure communication mechanisms without necessitating manual alterations to the services. By automating the signing process, MPKLink facilitates the adoption of a secure communication framework in established microservices architectures while preserving the performance benefits inherent in shared memory-based communication.

\section{Methodology}
A benchmark implementation has been developed to assess and compare various inter-process communication (IPC) techniques within a microservices architecture, specifically using microservices implemented in Rust. The experimental setup comprises two distinct microservices—one predominantly read-intensive and the other primarily write-intensive—deployed on a single host machine via Docker. The services interact through multiple IPC mechanisms, with the \texttt{serde} crate employed for serialization and deserialization of data structures, serving a role analogous to that of protocol buffers. The integration of Intel Memory Protection Keys (MPK) is facilitated through the \texttt{pkey\_mprotect} crate, while shared memory operations are managed via the \texttt{shared\_memory} crate. For alternative IPC methods, the \texttt{nix} crate is utilized for managing operating system pipes, and Unix Domain Sockets provided by the Rust standard library are also leveraged.

Several communication paradigms were explored during the IPC implementations, with specific attention given to their inherent characteristics. For operating system pipes, two separate pipes were established to facilitate unidirectional communication, given that this configuration offers a straightforward and reliable setup. Although bidirectional communication over pipes is feasible, it may introduce synchronization challenges; hence, unidirectional communication was adopted to simplify coordination. Named pipes, created via the \texttt{mkfifo} utility, were employed to facilitate this mode of data exchange. Another IPC method implemented was Unix Domain Sockets, with a single socket configured to support bidirectional communication. Despite the blocking nature of sockets, which can introduce delays by ensuring that data is completely received before processing, they also guarantee reliable data delivery.

Shared memory was investigated as an additional IPC mechanism. In this approach, two distinct memory regions were allocated since shared memory does not inherently provide blocking behavior—unlike sockets or pipes—and this may elevate the risk of race conditions. To address these concerns, a polling mechanism was implemented whereby the server continuously monitors the shared memory buffer for metadata signals indicating that the client is prepared to read or write. Initial attempts to implement the MPK configuration with the services operating as entirely separate processes via file-backed memory-mapped shared memory resulted in segmentation faults and presented significant challenges in managing memory without corruption or race conditions. Consequently, a transition was made to a thread-based approach, under the assumption that a master process is available to orchestrate the microservices as individual threads. This modification allowed the memory tagged with protection keys to be mapped within a single process, thereby reducing the likelihood of memory corruption.

To ensure reproducibility and clarity, our experimental setup was configured as follows. Each microservice was allocated [specific memory size] and deployed as a thread within a single Docker container using [Docker version/settings]. The benchmarking environment was established on an Intel Skylake-based server with 32 cores, as detailed in Section VII, and utilized [specify parameters such as thread count, synchronization intervals, and buffer sizes]. Furthermore, we measured synchronization overhead by recording the time taken to update memory protection keys across threads, which provided insights into the performance degradation observed for larger payloads.

The benchmarking task was structured around a distributed word count problem. In this scenario, Service 1 (acting as the client) reads data from a file, serializes a request, and transmits it to Service 2 (serving as the server). The server deserializes the request, computes the word count for the file, and sends the resulting count back to the client, which then outputs the result to the standard output. This task was employed to evaluate both the performance and reliability of the various IPC methods under diverse conditions. Challenges related to the synchronization of memory protection keys across services were examined, with particular emphasis on ensuring that the implemented protection mechanisms do not adversely affect communication performance. Although refinements to the system and the benchmark setup continue to be pursued, the current implementation yields valuable insights into the comparative strengths and weaknesses of the different IPC approaches.

\section{Results}
A series of microservices was created and benchmarked on a C6420 Cloudlab Node. The testing environment consisted of an Intel Skylake-based server featuring 32 cores and dual disks, powered by two 16-core Intel Xeon Gold 6142 CPUs operating at 2.6 GHz, accompanied by 384GB of ECC DDR4-2666 memory, two Seagate 1TB 7200 RPM 6G SATA hard drives, and a dual-port Intel X710 10Gbe NIC. For each set of microservices, test cases were generated involving word counts ranging from 100 to 100,000,000 words, and the duration required to transmit, compute, and receive a complete "count words" request was recorded.

\begin{figure}[h]
    \centering
    \includegraphics[width=0.3\textwidth]{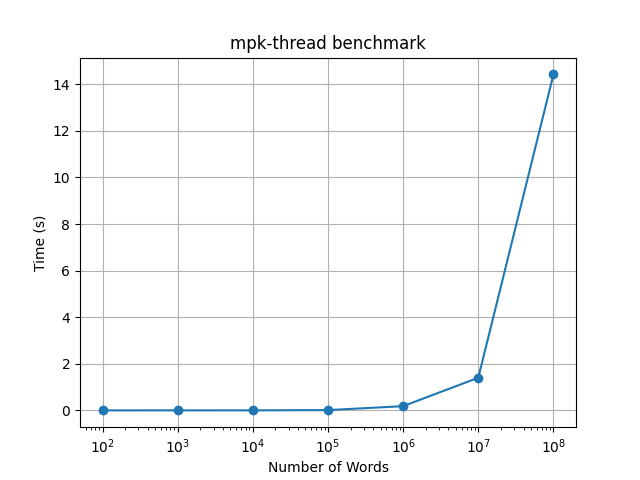}
    \caption{Benchmarking MPKLink}
\end{figure}

As illustrated in Figure 1, MPKLink demonstrates satisfactory performance for small-scale inter-service communications; however, an exponential increase in processing time is observed when handling larger payloads, particularly in the range between 10 million and 100 million words. This pronounced escalation in latency renders the current implementation of MPKLink impractical for managing large-scale bulk communications.

\begin{figure}[h]
    \centering
    \includegraphics[width=0.3\textwidth]{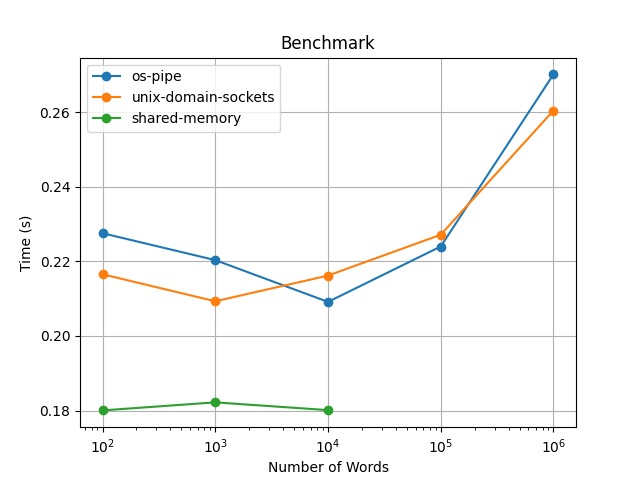}
    \caption{Overview of all results for all four IPC methods}
\end{figure}

Figure 2 presents a comparative overview of MPKLink alongside three alternative IPC communication methods. Although MPKLink achieves performance levels that are comparable to those of its counterparts, it does not exceed the speed of any other method across the full spectrum of word counts tested.

\begin{figure}[h]
    \centering
    \includegraphics[width=0.3\textwidth]{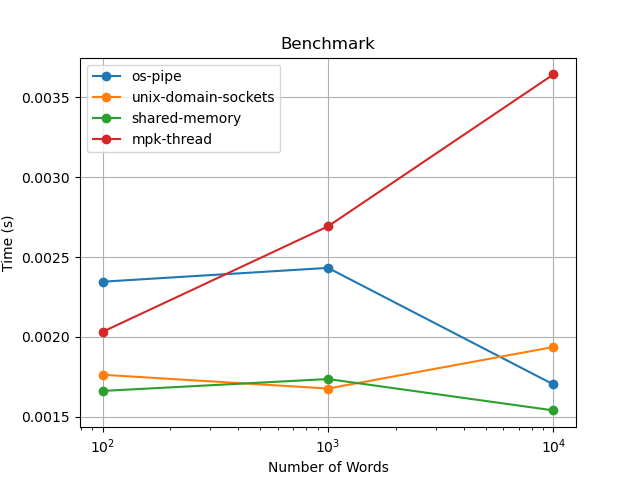}
    \caption{Overview of small results for all four IPC methods}
\end{figure}

In Figure 3, attention is specifically focused on smaller-scale inter-service communications (comprising 10,000 words or fewer). In this context, MPKLink outperforms operating system pipes by approximately 0.0005 seconds for communications involving 100 words or fewer, yet its performance remains inferior to that of shared memory and Unix Domain Sockets. Notwithstanding this, MPKLink provides enhanced isolation guarantees compared to shared memory, rendering it a viable option in scenarios where security and isolation are of paramount importance. Furthermore, the architecture of MPKLink—facilitating memory sharing between two services operating as individual threads via an mmap-based region—enables it to effectively manage larger communications. In contrast, the baseline shared memory approach is incapable of handling requests involving 100,000 words or more.

\begin{table}[h!]
    \centering
    \label{tab:ipc-best}
    \begin{tabular}{|r|l|l|}
        \toprule
        \hline
        \textbf{Word Count} & \textbf{MPKLink (s)} & \textbf{Best Other (s)} \\
        \midrule
        100       & 0.00203        & 0.00166 (Shared Memory) \\
        1,000     & 0.00269        & 0.00168 (Unix Sockets) \\
        10,000    & 0.00364        & 0.00154 (Shared Memory) \\
        100,000   & 0.01536        & 0.00660 (OS Pipe) \\
        1,000,000 & 0.18374        & 0.04571 (Unix Sockets) \\
        10,000,000 & 1.40530       & 0.48885 (Unix Sockets) \\
        100,000,000 & 14.42533     & 5.10027 (Unix Sockets) \\
        \hline
        \bottomrule
    \end{tabular}
    \caption{Comparison of MPKLink and Best-Performing IPC}
\end{table}

Finally, we provide a side-by-side comparison of MPKLink with the best-performing IPC method for each word count. While MPKLink shows competitive performance for small payloads, it is consistently slower than the best-performing alternatives, especially as the word count increases.

\section{Related Work}
A substantial body of research has focused on improving both the security and efficiency of inter- and intra-process communication in modern computing environments. Many early works in this area concentrated on using hardware features to enforce isolation within a single process. For instance, Hodor~\cite{hodor2019intra} demonstrated the application of Intel Memory Protection Keys (MPK) to achieve fine-grained intra-process isolation by dynamically modifying access rights, thereby protecting sensitive data during execution. Similarly, ERIM~\cite{vahldiek2019erim} integrated MPK with binary introspection and hardware watchpoints, allowing for secure, efficient in-process isolation with minimal performance overhead.

In parallel, several studies have examined traditional communication protocols within microservices architectures. Research such as the work by Doe and Smith~\cite{doe2020performance} has characterized the performance of communication protocols like REST and gRPC, identifying significant overheads when these methods are used for inter-microservice communication on the same host. Johnson and Lee~\cite{johnson2021serverless} further investigated the performance limitations in serverless computing environments, highlighting the latency introduced by conventional network protocols.

Shared memory has been recognized as a promising alternative for reducing communication latency. Zhang and Kumar~\cite{zhang2018optimizing} explored optimizations in shared memory IPC for multi-core systems, while Patel and Gupta~\cite{patel2019secure} proposed methods to secure data exchanges in containerized environments. However, these approaches often lacked the necessary fine-grained access control, which is critical when multiple processes or microservices share common memory resources.

To address these challenges, several researchers have turned to hybrid solutions. Fernandez and Martin~\cite{fernandez2020survey} provided a comprehensive survey of inter-process communication mechanisms in Linux, noting that while shared memory can offer high performance, it generally requires additional security measures to prevent unauthorized access. Chen and Huang~\cite{chen2020dynamic} investigated dynamic memory access control techniques in high-performance computing, emphasizing the need for rapid and secure permission modifications without incurring significant overhead.

Recent studies have also examined secure communication in microservices through cryptographic enhancements. Lee and Park~\cite{lee2021enhancing} presented a framework for enhancing microservice security with lightweight cryptography and digital signatures, ensuring the integrity and authenticity of inter-service communications. Singh and Rao~\cite{singh2022secure} discussed the broader challenges of securing microservice communication and proposed integrated solutions that combine secure IPC methods with modern orchestration platforms.

The current approach builds upon these diverse research efforts by leveraging Intel MPK not only for protecting data at rest but also for securing shared memory used in inter-microservice communication. By combining MPK-based access control with traditional shared memory techniques, the proposed framework addresses the performance limitations highlighted in earlier works while simultaneously providing robust security guarantees.

\subsection{Comparative Analysis with Security-Enhanced IPC Frameworks}
To contextualize the contributions of MPKLink, we conducted a comparative analysis with existing security-enhanced IPC frameworks. Unlike traditional frameworks that primarily focus on securing data at rest or rely on conventional access control mechanisms, MPKLink utilizes Intel MPK to dynamically manage memory permissions. This approach provides a dual advantage: it significantly reduces latency by avoiding costly system calls (e.g., mprotect) and enhances security by offering fine-grained, thread-local control. Table X summarizes the differences in performance metrics and security features, highlighting MPKLink’s strengths in environments where both low latency and stringent isolation are critical.

\section{Conclusion}
The present work introduced MPKLink, a novel framework that leverages Intel Memory Protection Keys (MPK) to enhance intra-container communication in microservice architectures. By integrating shared memory techniques with MPK-based dynamic access control, MPKLink addresses the inefficiencies inherent in traditional IPC methods such as REST and gRPC while simultaneously providing robust security guarantees. Experimental evaluations have shown that MPKLink delivers competitive performance for small-scale communications and offers strong isolation between microservices. However, the observed exponential increase in latency for larger payloads indicates that further optimizations are required to improve scalability in high-volume scenarios.
While MPKLink demonstrates competitive performance for small-scale inter-service communications, our results indicate an exponential increase in latency for large payloads. This degradation is primarily due to the overhead of synchronizing memory protection keys across threads. Future work will explore optimizations in the memory synchronization mechanisms and investigate hybrid IPC solutions that combine the low latency of shared memory with additional scalability.

In addition, although our security model is robust, a more detailed threat analysis is warranted. Future iterations of MPKLink should include a comprehensive examination of potential attack vectors—such as side-channel attacks, unauthorized memory access, and memory scraping—and incorporate mitigation strategies including advanced cryptographic safeguards, real-time anomaly detection, and enhanced access logging. Such measures will further fortify MPKLink’s suitability for real-world deployment in hostile environments.

Future work will focus on refining the synchronization mechanisms for memory protection keys and exploring hybrid communication approaches that can more effectively handle larger data transfers. In addition, the integration of MPKLink with modern container orchestration platforms will be further investigated to streamline its adoption in real-world environments. Overall, the contributions presented in this work provide a promising foundation for secure and efficient microservice communication, paving the way for subsequent advancements in inter-process communication technologies.

\end{document}